\documentclass{article}

\usepackage{PRIMEarxiv}

\usepackage[utf8]{inputenc} 
\usepackage[T1]{fontenc}    
\usepackage{hyperref}       
\usepackage{url}            
\usepackage{booktabs}       
\usepackage{amsfonts}       
\usepackage{nicefrac}       
\usepackage{microtype}      
\usepackage{lipsum}
\usepackage{fancyhdr}       
\usepackage{graphicx}       
\graphicspath{{media/}}     
\usepackage{siunitx}
\usepackage{soul}

\title{Self-Pulsing Microring Resonator Networks for Bandwidth-Efficient Event Detection in an Optical Fiber Sensor
}

\author{
  Alessio Lugnan \\
  Nanoscience Laboratory, Department of Physics \\
  University of Trento \\
  Trento, Italy\\
  \texttt{alessio.lugnan.1@unitn.it} \\
     \And
  Yonas Seifu Muanenda \\
  Institute of Mechanical Intelligence\\
  Scuola Superiore Sant'Anna\\
  Pisa, Italy\\
   \And
  Ilya Auslender \\
  Nanoscience Laboratory, Department of Physics \\
  University of Trento \\
  Trento, Italy\\
   \And
  Stefano Biasi \\
  Nanoscience Laboratory, Department of Physics \\
  University of Trento \\
  Trento, Italy\\
     \And
  Claudio J. Oton \\
  Institute of Mechanical Intelligence\\
  Scuola Superiore Sant'Anna\\
  Pisa, Italy\\
     \And
  Fabrizio Di Pasquale \\
  Institute of Mechanical Intelligence\\
  Scuola Superiore Sant'Anna\\
  Pisa, Italy\\
   \And
  Lorenzo Pavesi \\
  Nanoscience Laboratory, Department of Physics \\
  University of Trento \\
  Trento, Italy\\
}

\begin{document}
\maketitle

\begin{abstract}
The native processing of time-dependent signals from optical sensors by integrated photonic circuits can potentially bring significant advantages in terms of energy consumption, latency and processing power, as it allows skipping or reducing the use of fast digital electronics and  directly exploiting optical degrees of freedom and parallelism. However, due to a short memory, optical operations usually struggle to directly process optical signals with relatively slow (<\SI{}{\mega\hertz}) dynamics from optical sensors. In this work, we experimentally show that these limitations can be overcome by exploiting the self-pulsing dynamics in a microring resonator (MRR) network. In particular, we demonstrate that such dynamics can expand and retain information about perturbations sensed by a fiber sensor. This reduces the minimum sampling rate for the digitization of the sensor signal by at least one order of magnitude. The reduction is achieved by combining fiber sensing measurements at two different perturbation locations and frequencies with MRR network measurements at multiple output ports, input power levels and laser wavelengths. This work represents a first step in bridging time-dependent optical processing and optical sensing at sub-\SI{}{\micro\second} time scales.
\end{abstract}


\section{Introduction}

Photonic computing offers emerging alternatives and exclusive enhancements to traditional electronics, featuring high parallelism, minimal latency, and energy efficient data transportation and linear operations  \cite{xu2023integrated, stroev2023analog, bente2025potential, brunner2025roadmap}. 
For technologies relying on continuous streams of optical data, such as fiber optic sensors and networks thereof, deploying neuromorphic hardware directly in the optical domain is strategic in achieving efficient real time processing. \cite{meng2021integrated,xiao2024multimodal,ren2025near,zhou2025technology,tao2025nanosecond}. For example, processing the time-dependent output of optical sensors natively avoids the substantial latency, bandwidth bottlenecks, and power penalties introduced by standard optical-to-electrical conversions. However, experimental demonstrations are scarce since useful operations on sensed temporal data are often nonlinear and require memory of past inputs. Indeed, due to the inherent lack of direct photon-photon interactions, achieving these computational properties in photonic processors remains a significant hurdle\cite{foradori2026memory}.

 Silicon microring resonators (MRRs) provide a promising and naturally suited solution to these limitations. These simple, CMOS-compatible devices \cite{bogaerts2012silicon} are extensively used in both optical communication and sensing due to their ability to tightly confine light and significantly enhance light-matter interactions. Operating at telecommunication wavelengths (e.g., 1550 nm), MRRs exhibit strong nonlinearities driven by two-photon absorption (TPA) alongside the associated free-carrier absorption and dispersion (FCA and FCD) \cite{leuthold2010nonlinear, borghi2017nonlinear}. TPA generates free carriers within the waveguide that undergo thermalization, locally raising the temperature and modifying its refractive index via the thermo-optic (TO) effect. The MRR resonance frequency is subsequently subjected to a competing blue shift from FCD and a red shift from the TO effect \cite{johnson2006selfinduced, biasi2022effect}. More importantly, these optical effects possess distinct relaxation timescales: thermal lifetimes typically span 60–280 ns, whereas carrier lifetimes range from 1–45 ns \cite{vaer2012simplified, borghi2021modeling}. When driven by a continuous wave input near the resonance wavelength, this interplay of competing shifts and mismatched temporal dynamics forces the MRR into self-pulsing (SP) oscillations \cite{priem2005optical, pavesi2021thirty}. By coupling multiple silicon MRRs into a network architecture, the system can generate a rich variety of dynamic SP responses, including chaotic regimes, simply by tuning the input laser power and wavelength \cite{mancinelli2014chaotic}. This yields a highly nonlinear, memory-rich physical substrate that can provide a wide variety of responses to time dependent inputs, which can be exploited in native photonic processing of optical signals \cite{biasi2024photonic}.

Distributed Acoustic Sensing (DAS) transforms optical fibers into dense arrays of vibration sensors with the most common implementation utilizing phase-sensitive optical time-domain reflectometry ($\Phi$-OTDR) to detect dynamic strain via coherent Rayleigh backscattering \cite{muanenda2018recent}. Characterized by its long sensing range, high spatial resolution, high measurement bandwidth, \cite{hartog2017introduction}, DAS is widely employed in applications ranging from structural health and seismic monitoring to smart-city and oceanographic sensing \cite{fernandez2022seismic,qian2019distributed}. Recent advancements have focused on improving the signal-to-noise ratio and event classification through coherent detection \cite{martins2013coherent}, machine learning \cite{gemeinhardt2023machine}, and cloud-based architectures \cite{nur2024design} and those addressing polarization fading \cite{demise2025strategies}. However, to reduce the latency and high-speed digitization requirements inherent in digital processing, analog-domain front-end demodulation techniques, such as 3$\times$3 coupler-based retrieval and Phase-Generated Carrier (PGC) methods \cite{yu2018distributed} have also been proposed. Typical advanced signal processing in DAS focus on manipulation of the optical characteristics of the coherent light source and digital processing to obtain the single-pulse equivalent response with techniques including optical pulse coding and advanced pulse compresssion with nonlinear frequency modulation \cite{muanenda2022adaptable}. In parallel,nonlinear microring resonators (MRRs) have been extensively studied for analog optical signal processing, notably as artificial spiking neurons for neuromorphic computing \cite{van2012cascadable, xiang2022all, lugnan2022rigorous, biasi2024exploring, donati2025all, xiang2025photonic} or within reservoir computing frameworks \cite{mesaritakis2013micro, borghi2021reservoir,donati2024time,lugnan2025emergent,lugnan2025reservoir,foradori2025neuromorphic,donati2026photonic}. 

In this work, we bridge these two domains by demonstrating the application of a nonlinear MRR network to an optical fiber sensor. By exploiting the high sensitivity of SP dynamics to fast, nanosecond-scale input optical signals (a behavior previously investigated in \cite{biasi2024exploring,lugnan2025reservoir}) the network effectively amplifies and retains perturbation information over an expanded timescale. This novel photonic circuit-based approach reduces the digital processing and memory costs of DAS by lowering the required digitization sampling rate by at least an order of magnitude.

This article is structured as follows: We first explain the experimental measurements and data processing involved in this study and their rationale (Section \ref{subsec:application}). Subsequently, we discuss our results on the application of the MRR network to the detection of perturbation frequencies along the fiber sensor using a lower digitization sampling rate with respect to a standard approach (Section \ref{subsec:freq_detection}).  We then present an extension of these results, where we demonstrate similar advantages when also the perturbation location, in addition to the frequency, is recognized (Section \ref{subsec:loc_detection}). Finally, after summarizing and discussing our findings in the Conclusion section, in the Methods section we provide more technical details about measurements and data processing.

\subsection{Application of the MRR network to vibration sensing in an optical fiber}
\label{subsec:application}

The DAS fiber sensing data were obtained from multiple fiber measurements (more details are in Section \ref{methods_DAS}), each acquiring a matrix of coherent Rayleigh backscattering signals from 500 consecutive pulses. Each backscattering trace provides information on the (approximately) instantaneous state of the fiber. Instead, variations over subsequent backscattering traces capture the changes of fiber state over time, which are produced by a periodic oscillatory perturbation imparted by two piezoelectric actuators placed in the middle (\textit{position 1}) and at the end (\textit{position 2}) of the \SI{395}{\meter} long fiber under test (FUT). Through these actuators, 7 different perturbation types were applied, namely: [off,off], [off,\SI{1}{\kilo\hertz}], [off,\SI{2}{\kilo\hertz}], [\SI{1}{\kilo\hertz},off], [\SI{1}{\kilo\hertz},\SI{1}{\kilo\hertz}], [\SI{2}{\kilo\hertz},off], [\SI{2}{\kilo\hertz},\SI{2}{\kilo\hertz}], where the position within the square brackets indicates the actuator position ([\textit{position 1}, \textit{position 2}]) and the values represent the frequency of the oscillatory perturbation. Each measurement of 500 traces was characterized by one perturbation type. In total, 5 repetitions of a sequence of 7 measurements (one for each perturbation type, in the same order as they are listed) were performed, amounting to 35 measurements. This intertwined chronological layout, as opposed to simply taking 7 long measurements, was employed to avoid data bias due to correlations between variations in measurement conditions and perturbation type. 

\begin{figure*}[t!]
	\centering
	\includegraphics[width=0.8\textwidth]{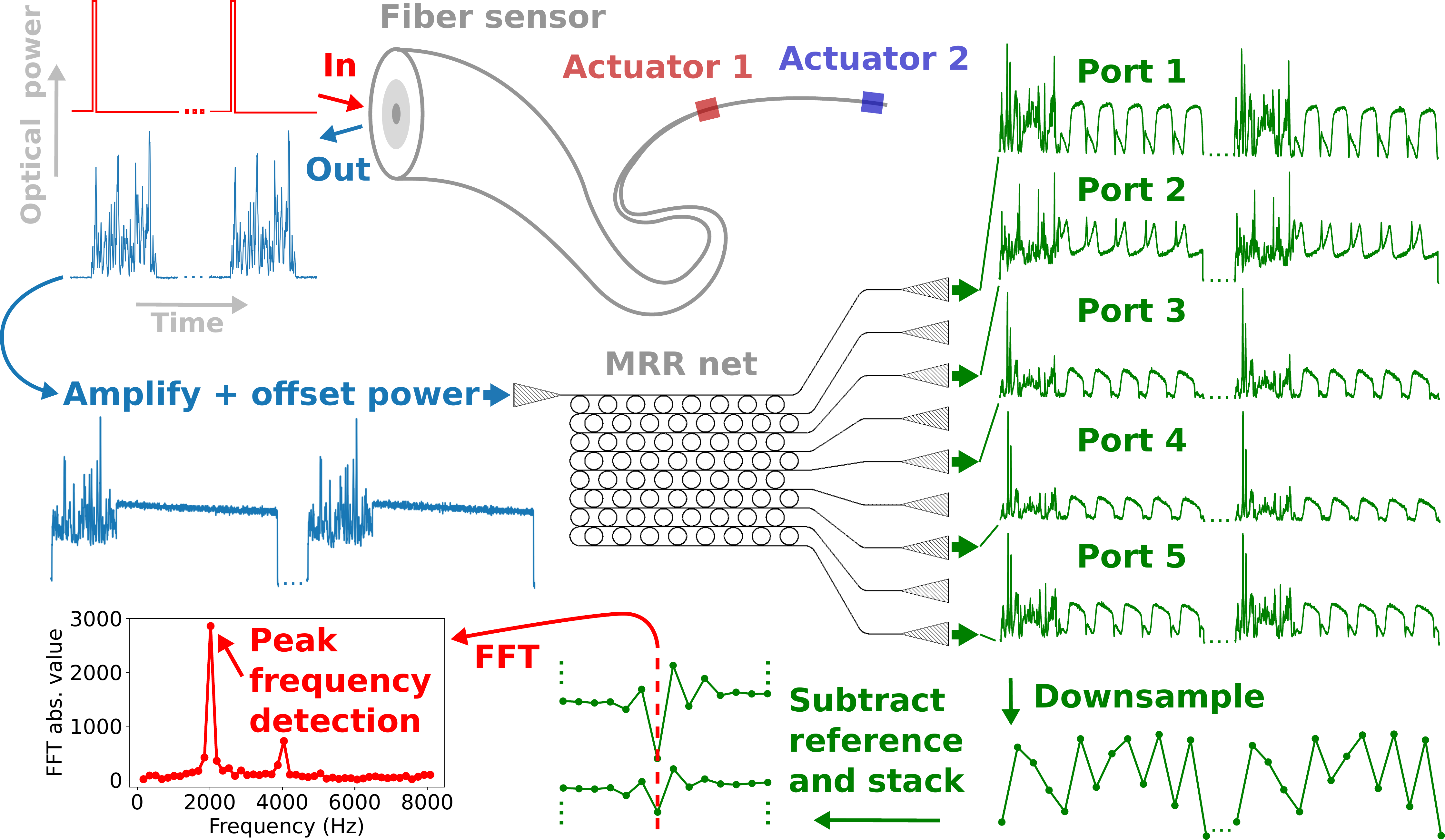}
	\caption{\textbf{Sensing and signal processing pipeline.} A \SI{395}{\meter} long DAS fiber sensor is perturbed by two piezoelectric actuators placed in the middle and at the end of the fiber (more details are in Section \ref{methods_DAS}). The sensor's output signal (i.e., the reflections of \SI{100}{\nano\second} input laser pulses at around \SI{1550}{\nano\meter} wavelength) is amplified, and an offset optical power (to sustain the MRR network dynamics) and low-power breaks (to reset the MRR network memory) are added. The resulting optical signal is then inserted into a MRR network. The network is described in Section \ref{sec:methods} and comprises \(8\times 8\) coupled MRRs with an input port and 8 outtput ports. The reset pauses enable the network to process one fiber pulse reflection after another, independently. Then, multiple nonlinear representations of the input signal are acquired for different network parameters (namely output port, input laser wavelengths and power levels). The digitized signals are then down-sampled (low-pass filter + linear interpolation) to approximate the use of slower photodetectors and lower sampling rate. Afterwards, conventional data processing is applied to the processed data to detect the frequency of the fiber perturbations. A reference sample (the first in the corresponding sequence) is subtracted to each sample. The samples from a sequence are then stacked into a real-valued matrix (each sample is a row, each time sample is a column) and the FFT is applied to each column. The detected frequency at each time sample is the one with highest FFT absolute value.}
 \label{fig:pipeline}
 \end{figure*}

It should be stressed that, due to the high sensitivity of the backscattering from the fiber to environmental perturbation, each measurement is characterized by a very different reflection signal (Figure \ref{fig:DAS_setupNprofiles} (b)). To overcome this issue, in conventional DAS data processing a reference backscattering trace is digitally subtracted from all other traces in order to eliminate the information about the slow varying reflection signal background. Such a background variability considerably complicates the tasks demonstrated in this work, since a clean background subtraction is not possible due to the nonlinearity of the signal transformation effected by the MRR network \cite{lugnan2025reservoir}.

In order to demonstrate DAS operations with low sampling rate, we implemented the signal processing pipeline described by Figure \ref{fig:pipeline}. The pulse reflections from the DAS fiber (detailed in Section  \ref{methods_DAS}) were detected, digitally preprocessed and used to modulated an optical signal from a CW laser which was then injected into the MRR network (details are given in Section \ref{methods_MRRnetMeas}). This methodology emulates a purely optical connection while conditioning the input optical signal to the MRR array. In fact, the different pulse reflection waveforms were shuffled and distanced by a break period of \SI{10}{\micro\second} with low (close to zero) optical power. Moreover, an offset power of 40\% of the original dynamic range was added to each pulse reflection and to the \SI{10}{\micro\second} right after, forming a plateau to excite and read the nonlinear response (e.g. self-pulsing) of the MRR network (examples are shown in Figures \ref{fig:pipeline} and \ref{fig:setup} (b)). The role of the low power break, instead, is to reset the optical memory of the MRR network, so that each sample (comprising the pulse reflection and the constant power plateau) is independent of the previous one. It should be stressed that this preprocessing is compatible with an all-optical implementation where the DAS fiber is optically connected to the MRR network. In such a case, the DAS fiber output would be extracted using a circulator, optically amplified (e.g., by an EDFA), and the constant power plateaus and breaks for pumping the network would be conveyed by a different wavelength with respect to the one used to probe the fiber. This approach removes the optical phase information in the reproduced pulse reflections, but phase variations usually occur on timescales much longer than the transit time of optical signals in the PIC (tens of picoseconds), thus we expect them to be negligible in this case.

Going through the processing pipeline, the signals coming out of the MRR network, which are obtained for different network parameters (i.e., output ports, laser wavelengths and power levels), are down-sampled to diverse sampling rates (details are provided in Section \ref{methods_downsamp}). Afterwards, we subtract a reference reflection signal (the first one in the corresponding DAS measurement) from all the other signals in the measurement, in order to reduce the influence of the slowly varying DAS background profiles, as it is done in conventional DAS processing \cite{Demiseetal2023}. Finally, a fast Frourier transform (FFT, absolute value) is applied to each sample point, obtaining the frequency spectrum of the reflection variations at a fixed sampling time. From this, the perturbation frequency is extracted by finding the frequency corresponding to the highest amplitude in the spectrum (peak frequency detection).

\section{Experimental Results}
\label{sec:application}

\subsection{Frequency detection at low sampling rates}
\label{subsec:freq_detection}

\begin{figure*}[t!]
	\centering
	\includegraphics[width=1.0\textwidth]{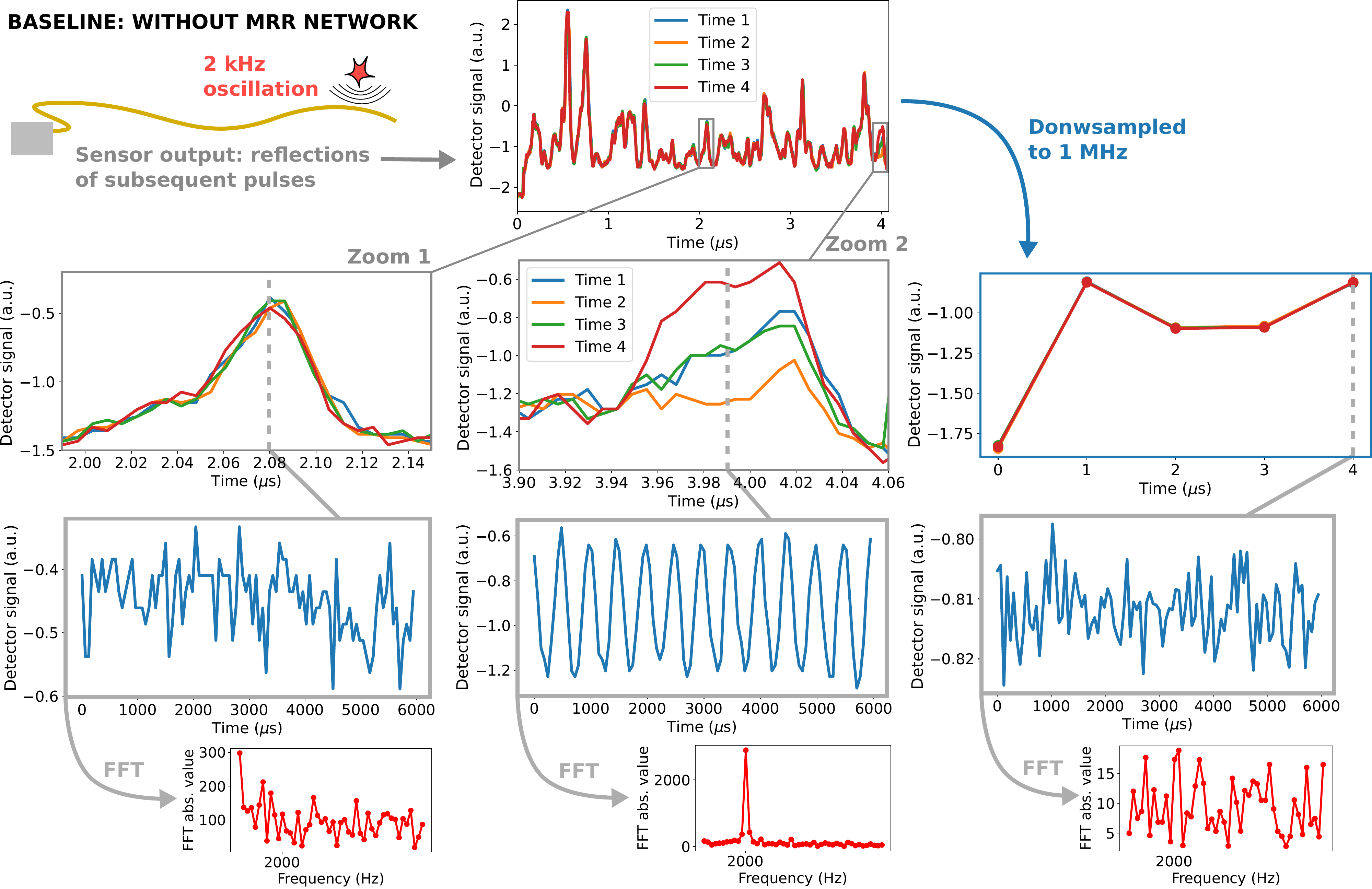}
	\caption{\textbf{Frequency detection without MRR network (baseline) - example plots.} In this example the FUT is perturbed with a \SI{2}{\kilo\hertz} oscillation at position 2 (end of the fiber). As expected in DAS measurements, the perturbation results in an oscillation of the pulse reflections at a specific time interval corresponding to the time the pulse takes to arrive at and propagate back from the location of the perturbation along the fiber. Indeed, Zoom 1 (first plot column) shows no oscillation at timings corresponding to position 1. This is also confirmed by the temporal behavior of the signal recorded at a fixed time indicated by a gray dashed line in the Zoom 1. The absence of periodic variations can be clearly observed. In addition, this is also evidenced by the absence of a clear peak in the FFT spectrum at a fixed reflection sampling time for subsequent reflections. Instead, Zoom 2 (second plot column) shows significant oscillations in reflection amplitude at the timing corresponding to position 2 of the perturbation. Oscillations are observed in the different colored curves for different measurement times. A periodic oscillation is also measured at the properly set fixed time (dashed line and bottom plot). In this case, the FFT spectrum of this signal presents a clear peak at the perturbation frequency. Finally, down-sampling the reflections signals to \SI{1}{\mega\hertz} (third plot column) prevents the detection of the perturbation.}
 \label{fig:examples_baseline}
 \end{figure*}

In order to properly evaluate the advantages of employing a MRR network for DAS, we first applied conventional frequency detection based on FFT to the reflection signals, thus obtaining baseline results without the use of the MRR network. Figure \ref{fig:examples_baseline} shows examples of the baseline frequency detection procedure applied to: (i) a time interval not corresponding to the perturbation location, (ii) a time interval corresponding to the timing corresponding to the perturbation location and, (iii) the same timing but employing a down-sampled reflection signal (from \SI{200}{\mega\hertz} to \SI{1}{\mega\hertz}). This last setting simulates the usage of a slow photodetector (1 MHz bandwidth) and 1 MSa/s digitization (details in Section \ref{methods_downsamp}). From the corresponding FFT spectrum it can be noticed that, in this case, the ability to detect the perturbation frequency is completely lost due to the low sampling rate. This is because each time sample of the down-sampled reflections integrates the time variations in the original reflection signals over a time interval that is much larger than one corresponding to the fiber perturbation.

\begin{figure*}[t!]
	\centering
	\includegraphics[width=1.0\textwidth]{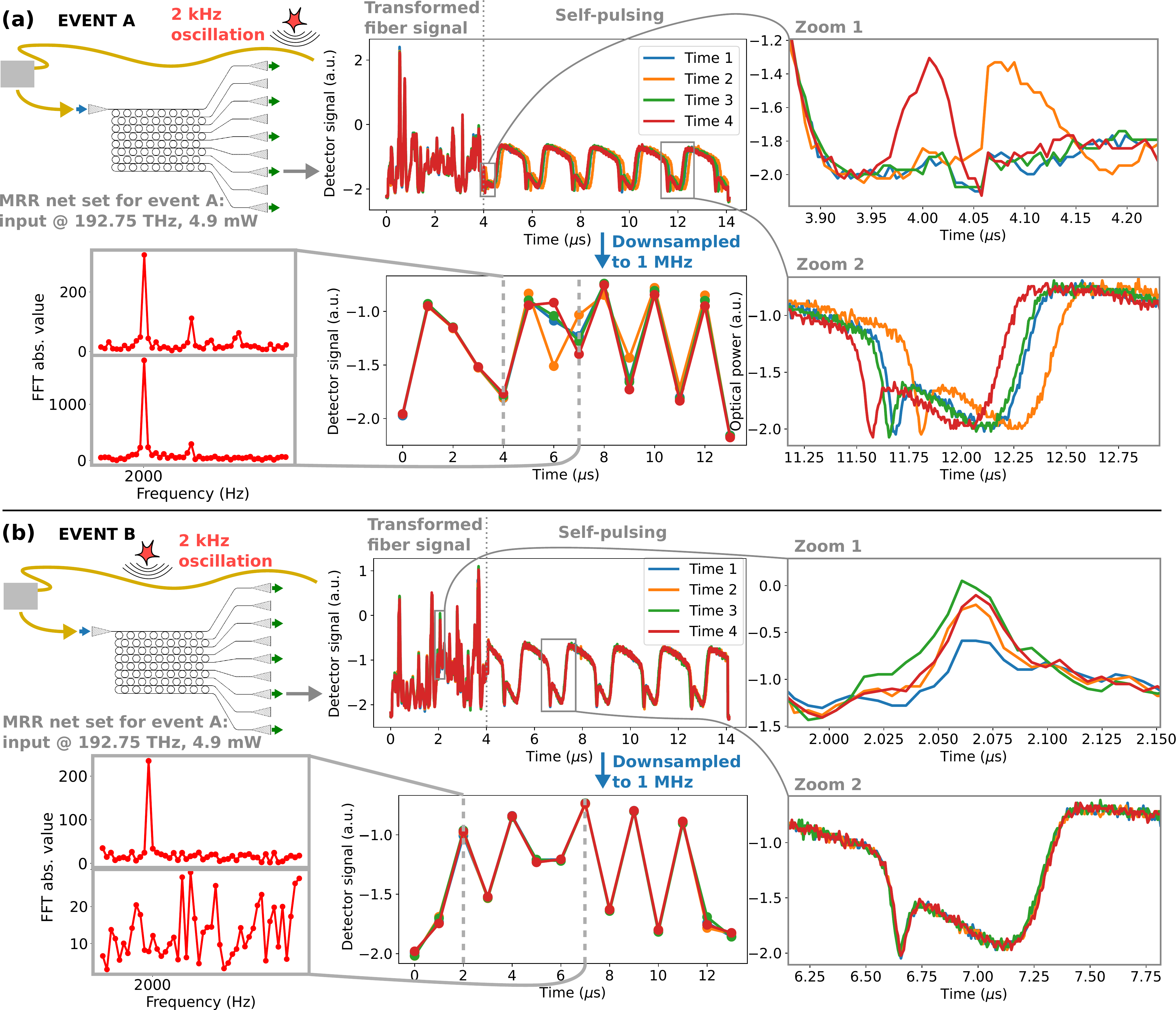}
	\caption{\textbf{Frequency detection using the MRR network - example plots.} Similarly to Figure \ref{fig:examples_baseline}, these plots show how the backscattering signal from the fiber is represented by the self-pulsing MRR network and the process we use to extract the perturbation frequency. \textbf{(a)} Response to an \textit{event A} (\SI{2}{\kilo\hertz} perturbation at position 2) and the corresponding FFT analysis after down-sampling the output signal to \SI{1}{\mega\hertz}. It can be noticed that the self-pulsing dynamics of the MRR network enable the detection of the perturbation frequency even after down-sampling, which was not possible in the baseline example represented in Figure \ref{fig:examples_baseline}. \textbf{(b)} Response to an \textit{event B} (\SI{2}{\kilo\hertz} perturbation at position 1) and the corresponding FFT analysis after down-sampling the output signal to \SI{1}{\mega\hertz}. In this case, it is not possible to detect the perturbation frequency from the down-sampled self-pulsing response. This shows that the MRR network dynamics can be selective regarding the location of the fiber perturbation, since, in this example, they enable frequency detection only at the location 2. An overview of the results obtained by exploiting these effects are provided in Figures \ref{fig:results_freq} and \ref{fig:results_loc}.}
 \label{fig:examples_MRRnet}
 \end{figure*}

The same procedure was then applied to the signal processed by the MRR array (Figure \ref{fig:examples_MRRnet}). In this case, the use of the self-pulsing MRR network allows to extract the perturbation frequency also from the down-sampled signal. This is enabled by the network's self-pulsing response, which amplifies and temporally extends the influence of the fiber perturbation on the optical signals. Figure \ref{fig:examples_MRRnet} (a) Zoom 1 and 2 show that both the processed reflected signal (Zoom 1) and the SP signal (Zoom 2) are affected by the perturbation. In the latter case, the perturbation affects the different SP periods due to the MRR intrinsic non-fading memory \cite{lugnan2025reservoir}.  Therefore, the down-sampled signals both for a time corresponding to the end of the reflected pulse (4 $\mu$s) as well as the one for a subsequent, arbitrary time instant (7 $\mu$s in this example) retain the information about the perturbation. An FTT on the temporal signal at these fixed time instances allows to extract the proper frequency, where a considerably greater signal-to-noise-ratio is achieved in the SP case.

We systematically repeated this processing for all the measurements for the different cases described in Section \ref{subsec:application}). We then calculated a detection error by taking the absolute value of the difference (in Hz) between the detected frequency and the actual perturbation frequency (ground-truth). We average these error values to obtain an \textit{average error}. This was done for different sampling frequencies, namely 13 values spanning three orders of magnitude from \SI{0.1}{\mega\hertz} to \SI{100}{\mega\hertz}, evenly spaced in the logarithmic scale (extremes included).

\begin{figure*}[t!]
	\centering
	\includegraphics[width=0.8\textwidth]{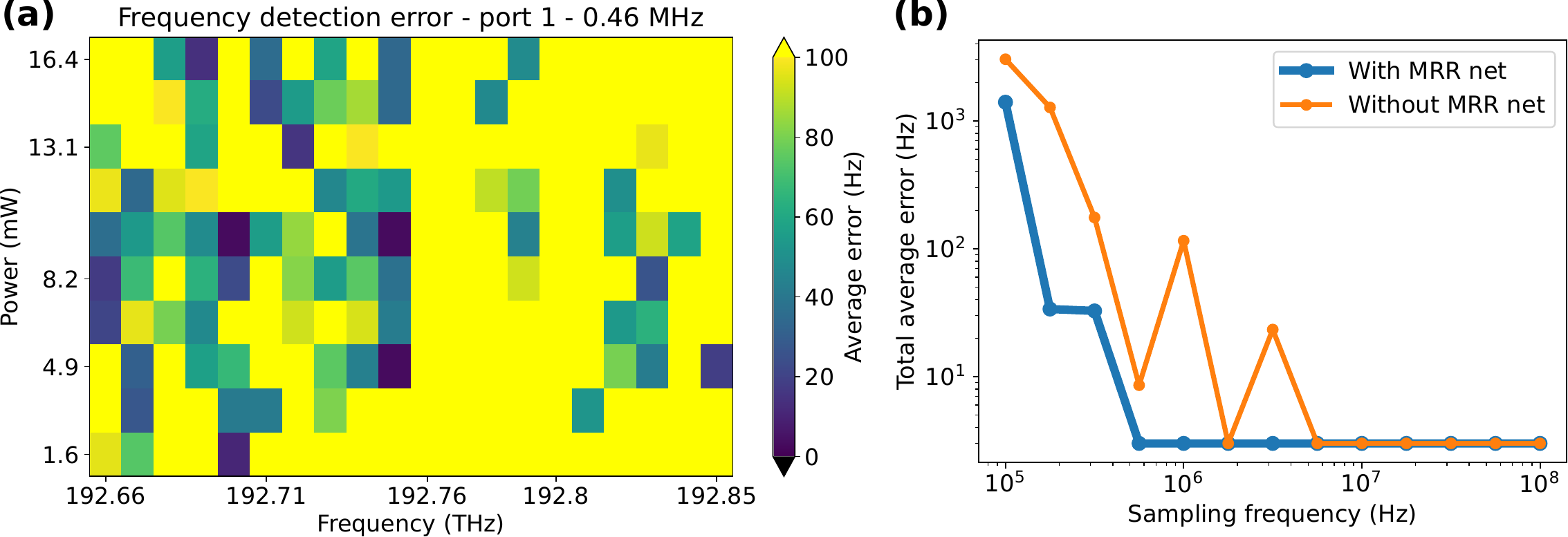}
	\caption{\textbf{Frequency detection error.} \textbf{(a)} Exemplary colormap of the average (frequency detection) error as a function of the input laser frequency and (on-chip average) power, at output port 1 and with \SI{0.46}{\mega\hertz} sampling frequency. Most points in the map present high error (yellow), but a few can detect the perturbation frequency with low error (dark blue). \textbf{(b)} Total average error (the best over the MRR network parameters) as a function of the sampling frequency. The MRR network (blue points) enables correct frequency detection down to a sub-MHz sampling frequency range, while the baseline results (orange points) presents a high error up to a sampling rate of \SI{1}{\mega\hertz}, and some significant error also at \SI{4.6}{\mega\hertz}.}
 \label{fig:results_freq}
 \end{figure*}

The average error values depend on the employed sampling rate, but also on the network control parameters (i.e., output port, laser wavelength and power). Indeed, some combinations of parameters, corresponding to different self-pulsing dynamics in the network, are more sensitive to the perturbation frequency than others. As an example, Figure \ref{fig:results_freq} (a) shows the colormap of the average frequency detection error as a function of laser frequency and power, using the output signals at port 1 and with a sample rate of \SI{0.46}{\mega\hertz}. It can be noticed that only a few points in the colormap present accurate frequency detection (dark blue points, error close to zero). 

A \textit{total average detection error} estimate is computed by selecting the best result for various network parameters for each sampling rate value. In this way,  an evaluation of the overall detection capability of the MRR network processing can be obtained. The use of the MRR network processing enables low frequency detection errors for a wide range of sampling rate, down to the sub-MHz range (namely down to \SI{0.56}{\mega\hertz}, see Figure \ref{fig:results_freq} (b)).  In comparison, the baseline results presents reliable detection only for sampling rates higher than \SI{5.6}{\mega\hertz}.
This factor-10 improvement in minimum sampling rate enables cheaper electronic digital processing for DAS and a corresponding reduction in memory usage and computational power. These advantages are particularly relevant in situations where multiple DAS fiber sensors are employed in a distributed sensing network.

 \subsection{Frequency and location detection at low sampling rates}
\label{subsec:loc_detection}
Conventional DAS allows the simultaneous detection of the frequency and location of a perturbation. On the other hand, we have shown that self-pulsing MRR network spreads the information about the perturbation frequency over longer time intervals, thus breaking the one-to-one correspondence between location along the fiber and timing of arrival of the reflected pulse. One might wonder whether also the location information can be recovered  with the processing of the reflected signal by a MRR network.

\begin{figure*}[t!]
	\centering
	\includegraphics[width=0.7\textwidth]{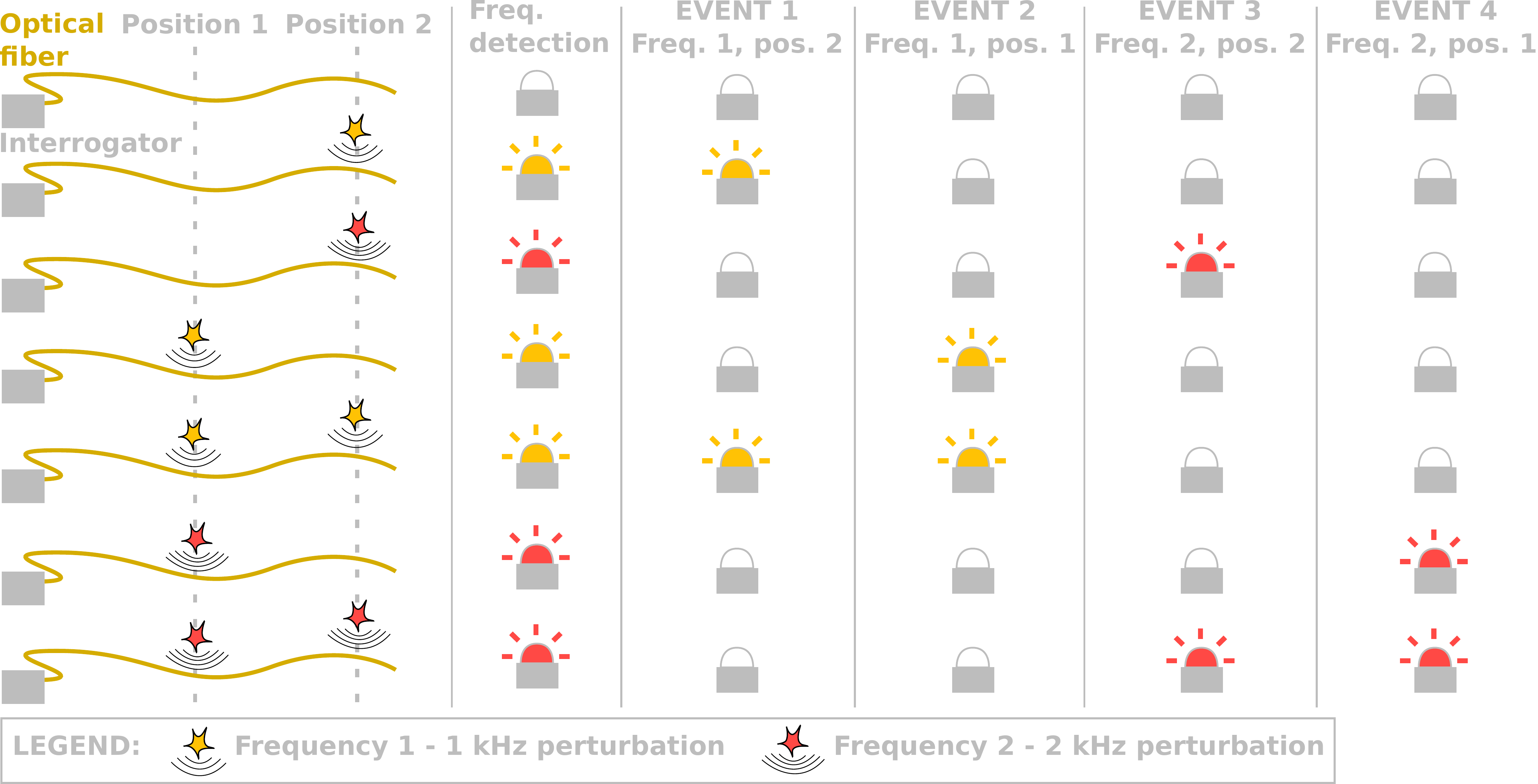}
	\caption{\textbf{Perturbation detection events.} In the first column, the 7 perturbations types are schematized (a reference and combinations of two perturbation frequencies and 2 locations). The second column shows the target output for the frequency detection task (Section \ref{subsec:freq_detection}).  The other 4 columns represent the other 4 detection tasks with different target event (specific frequency and location). Symbols: yellow light = \SI{1}{\kilo\hertz} perturbation is detected, orange light = \SI{2}{\kilo\hertz} detected.}
 \label{fig:locNfreq_explanation}
 \end{figure*}

Let us split the available measurements into 5 sub-measurements, that are disjointed sequences of 100 pulse reflections. This way, we improve the statistics by increasing the number of samples. Now we consider 4 different event detection tasks, each consisting in a different grouping of the sub-measurements in positive set of perturbations (that have to be detected) and negative set of perturbations (that should not be detected). The positive set corresponds to a target combination of perturbation frequency and location; negative set comprises all the other sub-measurements. In particular, the first event detection task targets perturbations with 1 kHz frequencies at location 1 (in the middle of the FUT). Thus, the positive set contains pertubations [1 kHz, off] and [1 kHz, 1 kHz] (notation introduced in Section \ref{subsec:application}), while the others belong to the negative set. Similarly, the other 3 event detection tasks respectively labels as positive set the following pairs of perturbations: [off, 1 kHz] and [1 kHz, 1 KHz]; [2 kHz, off] and [2 kHz, 2 kHz]: [off, 2 kHz] and [2 kHz, 2 kHz]. A visualization of this grouping of measurements into events is given by Figure \ref{fig:locNfreq_explanation}. Therefore, each event detection task is composed by 150 reflection sequences (each one constituted by 100 pulse reflections): 50 sequences are to be detected, 100 are to be ignored.

The event detection mechanism we employ and evaluate here is the following: given a target event, the same frequency detection procedure described in \ref{subsec:freq_detection}) is applied; then, only if the target frequency is among the detected frequencies, then the target event (target frequency and target location) is detected. Note that if the target frequency is only applied at the wrong location, it should not be detected. This is possible because the MRR network parameters are chosen so that the system is only sensitive to perturbations at the target location, as we explain in the following paragraphs.

\begin{figure*}[t!]
	\centering
	\includegraphics[width=0.99\textwidth]{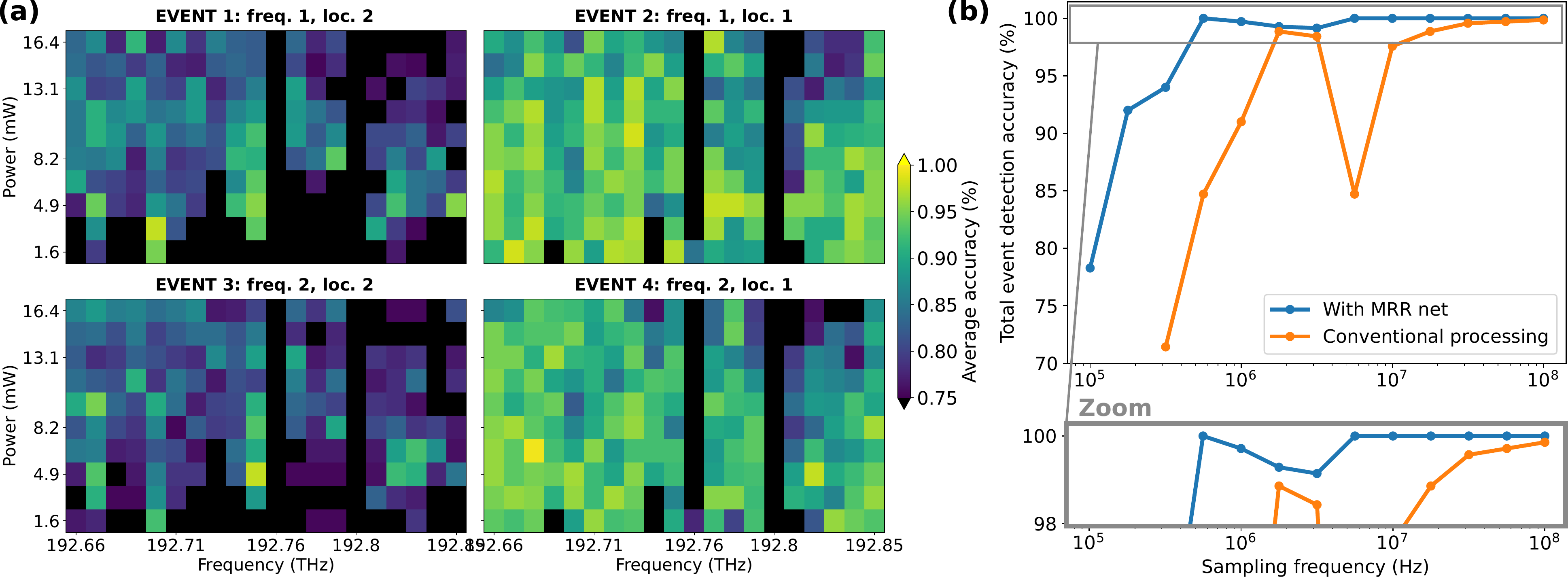}
	\caption{\textbf{Event (frequency and location) detection accuracy.} \textbf{(a)} Color-maps showing the detection average accuracy as a function of the MRR network control parameters, using output port 1, of the four events described in Figure \ref{fig:locNfreq_explanation}. Average accuracy color code is given on the left of the maps. \textbf{(b)} Total event detection accuracy obtained with optimal network parameters and timing, using the self-pulsing MRR network (blue) or the baseline method (orange, standard processing used as a baseline). It can be noticed that the MRR network enables: higher accuracy in general, lower sampling rate and more stable performances under sampling rate variations.}
 \label{fig:results_loc}
 \end{figure*}

We find that different network parameter combinations, and the associated self-pulsing dynamics, allow carrying out the 4 different event detection tasks with high accuracy for a wide range of sampling rates. The accuracy is calculated as one minus the average error rate. An example is provided in Figure \ref{fig:results_loc} (a), where an accuracy color-map for each event detection task is shown. These results refer to output port 1 and a sampling rate of \SI{0.46}{\mega\hertz}. It can be noticed that distinct points in the map perform very differently depending on the target event. Moreover, there is an overall correlation between the accuracy color-maps regarding event pairs sharing the same perturbation position, i.e., events 1 and 3, and events 2 and 4.

By selecting the best MRR network’s measurement parameters and the best timing (corresponding to a fixed time sample) in the digitized reflection signal for each event detection task, we obtained high event detection accuracy (above 99\%) for sampling rates down to \SI{0.56}{\mega\hertz} (Figure \ref{fig:results_loc} (b)). The accuracy raises to 100\% for sampling rates $>$ \SI{5.6}{\mega\hertz}. On the other hand, the baseline processing is always worse at a fixed sampling rate and achieves 100 \% accuracy only at \SI{200}{\mega\hertz} sampling rate. Notably, low accuracy (lower than 85\%) is found already at \SI{5.6}{\mega\hertz}. These results demonstrate that a self-pulsing MRR network allows monitoring DAS events (both frequency and location) with at least 10 times lower sampling rates than that required by a conventional approach. In a real application, the MRR network could be used to monitor the occurrence of target events at a low computational and memory cost (and using cheaper equipment) and, if a target event is detected, the more costly conventional DAS processing can be applied for target-independent analysis.

\section{Conclusion}
We experimentally demonstrate the interfacing of photonic processing by a self-pulsing microring resonator (MRR) network with a distributed acoustic sensing (DAS) system. We performed DAS measurements where the fiber was perturbed at two different locations, by vibrations at two different frequencies (\SI{1}{\kilo\hertz} and \SI{2}{\kilo\hertz}), and combinations thereof. We show that the self-pulsing dynamics in a MRR network allows us to amplify the DAS variations in pulse reflections, in terms of both amplitude and duration. This effect can be directly combined with conventional DAS data processing, with the advantage of reducing the required sampling rate for the optical signal digitization. This provides technological advantages, allowing the usage of cheaper instrumentation (slower photodetectors and digital-to-analog conversion (DAC) devices) and reducing the memory and computational cost required to operate a fiber sensor. Ultimately this will facilitate the employment of a large number of DAS devices with limited resources. 

However, these advantages come with a price: our method only allows monitoring perturbations at a target location at a time, which can be changed by modifying the measurement parameters. For example, multiple locations can be monitored concurrently by using multiple MRR networks, or using a single MRR network in different time slots corresponding to different segments of the sensing fiber. Nevertheless, these limitations can be potentially overcome by employing machine learning (accompanied by extensive DAS measurements to create a suitable training dataset) and/or by increasing the degrees of freedom available to control (or train) the MRR network’s self-pulsing dynamics, e.g. by modifying the MRR resonances via pn junctions \cite{biasi2024photonic}.

Finally, this work shows that photonic integrated circuits can be used for time-dependent processing of relatively slow optical signals from optical sensors, which is technologically advantageous but challenging, because of the difficulty to achieve efficient optical operations with volatile long memory (e.g., longer than \SI{1}{\micro\second}).

\section{Methods}
\label{sec:methods}

\subsection{DAS measurements}
\label{methods_DAS}

\begin{figure*}[t!]
	\centering
	\includegraphics[width=0.80\textwidth]{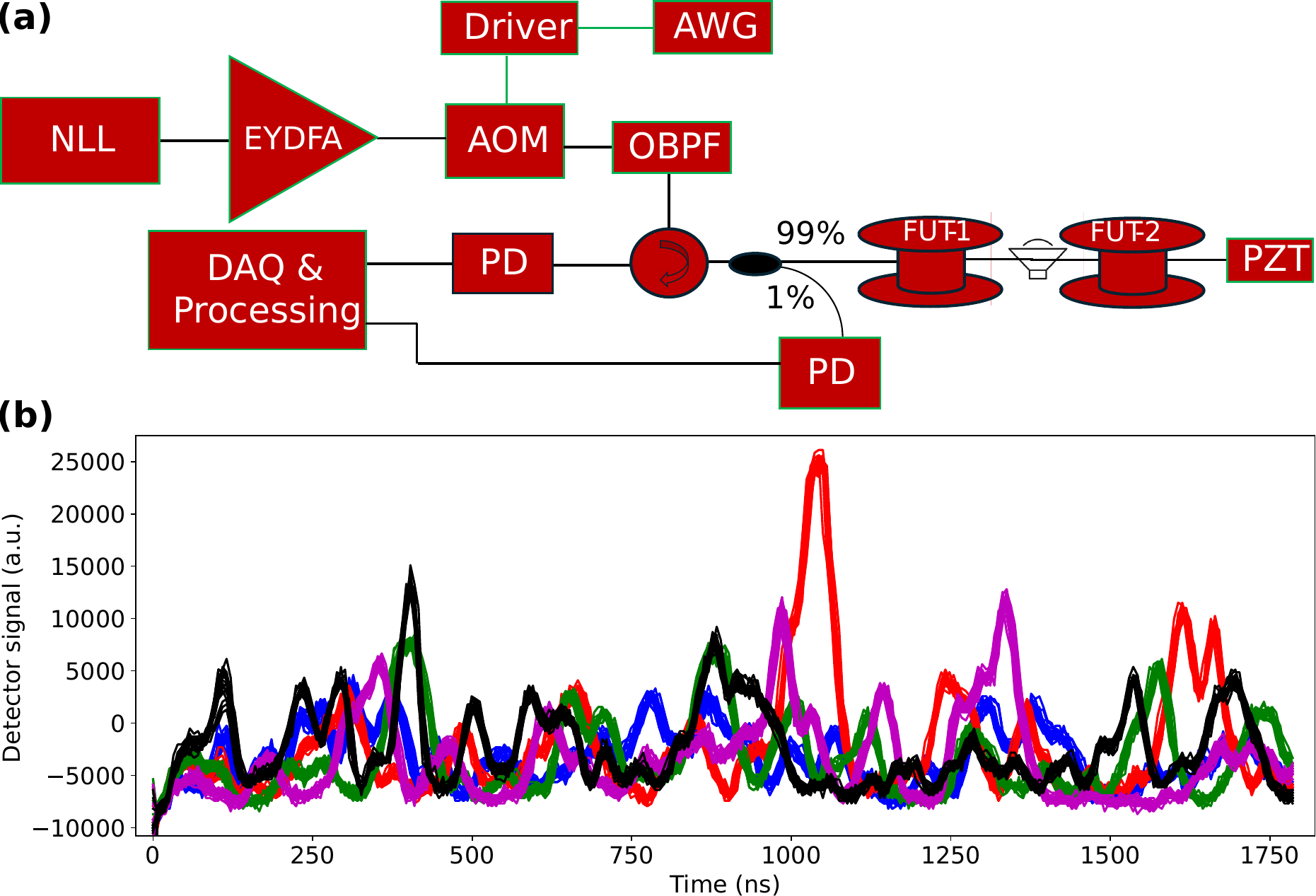}
	\caption{\textbf{DAS experimental setup and background variations over multiple measurement repetitions.} \textbf{(a)} Schematic of the experimental setup of reservoir computing with a DAS based on \(\Phi\)-OTDR. (NLL: Narrow Linewidth Laser; EDFA Erbium-Doped Fiber Amplifier; OBPF: Optical Bandpass Filter; AOM: Acousto-Optic Modulator; AWG: Arbitrary Waveform Generator; DAQ: Digital Acquisition; FUT: Fiber Under Test, PD: Photodiode; PZT: Piezoelectric Actuator.) \textbf{(b)} Here we plot the acquired coherent Rayleigh backscattering signals at the output of the FUT for 5 different repetitions (corresponding to different data colors) of the same measurement type, characterized by no oscillations applied by the piezoelectric actuators. It can be noticed that the reflection signal background, while it is relatively stable within the same measurement (same data color), undergoes considerable variations from one measurement to another. By comparing these background variations with the signal variations due to the fiber perturbations to be sensed (see example in Figure \ref{fig:examples_baseline}), we notice that the first are far larger than the latter. This significantly complicates analog signal processing for DAS, as it makes background subtraction difficult to avoid.}
\label{fig:DAS_setupNprofiles}
\end{figure*}

The experimental setup of the DAS sensor based on phase-sensitive optical time-domain reflectometry (\(\Phi\)-OTDR) used to measure multiple vibrations is depicted in Figure \ref{fig:DAS_setupNprofiles} (a). As shown, first the signal from a highly coherent, narrow-band laser with a linewidth of 100 Hz is amplified using an Erbium-Ytterbium-Doped Fiber Amplifier (EYDFA), and the ASE noise is suppressed with an Optical Bandpass Filter (OBPF). Subsequently, the probing optical pulses are generated using a high-extinction ratio AOM, which further suppresses the remaining ASE in the zero level, with a dedicated driver fed with controlled RF pulses from the Arbitrary Waveform Generator (AWG). 
A 1\% tap of the signal is photodetected to be used to both monitor the probe power, pulse width, and repetition rate and to trigger the acquisition when real-time measurements of vibration are performed to verify the consistency between applied and measured perturbations, so as to suppress  jitter affecting the time correspondence of subsequent pulses and traces. The 99\% tap is sent to the Fiber Under Test (FUT) through a three-port circulator.

The coherent Rayleigh backscattering signal at the circulator’s return port is then detected using a simple PIN photodiode of \SI{125}{\mega\hertz} bandwidth, and acquired using a DAQ with a sampling rate of 200 MS/s. Vibration testing was conducted by applying perturbations to short fiber segments at the middle and end of the FUT using a speaker and a PZT. These actuators are positioned at approximately 195 m and 395 m along the fiber, respectively. The system operates with a pulse width of \SI{100}{\nano\second}, corresponding to a spatial resolution of approximately 10 m, and a pulse repetition period of \SI{60}{\micro\second}. To simulate uncorrelated sources of perturbation, the two actuators are driven by sinusoidal signals generated by the AWG and a sound speaker at discrete frequencies of 1 kHz or 2 kHz, enabling both isolated and simultaneous perturbations. Seven actuation scenarios are selected from combinations of these excitation states to represent single-point and dual-point vibration conditions, resulting in a total of 35 vectors of raw coherent Rayleigh backscattering traces. Each configuration consists of five measurement rounds, with 500 \(\Phi\)-OTDR traces acquired per round. These configurations are summarized in Table \ref{tab:DASmeas}. It should be stressed that backscattering traces acquired at different measurement rounds present considerably different static background profiles (see Figure \ref{fig:DAS_setupNprofiles} (b)).

\begin{table}[h]
    \centering
    \caption{Applied vibration combinations.}
    \label{tab:DASmeas}
    \begin{tabular}{|c|c|}
        \hline
        Vibration 1 ($\approx$195 m, Speaker) & Vibration 2 ($\approx$395 m, PZT) \\
        \hline
        off   & off   \\
        \hline
        off   & 1 kHz  \\
        \hline
        off   & 2 kHz  \\
        \hline
        1 kHz  & off   \\
        \hline
        1 kHz  & 1 kHz  \\
        \hline
        2 kHz  & off   \\
        \hline
        2 kHz  & 2 kHz  \\
        \hline
    \end{tabular}
\end{table}

\subsection{MRR network measurements and experimental setup}
\label{methods_MRRnetMeas}

\begin{figure*}[t!]
	\centering
	\includegraphics[width=0.80\textwidth]{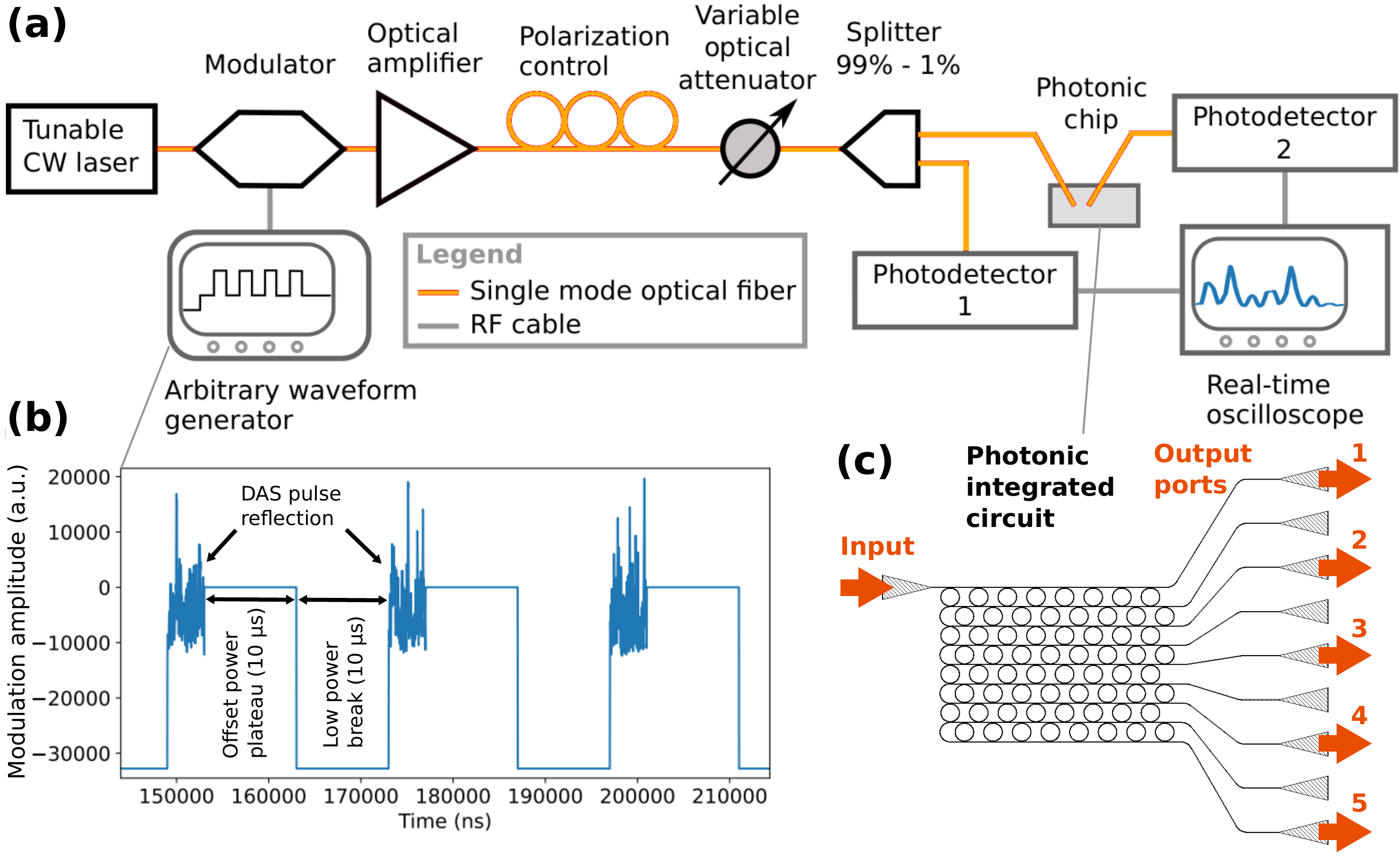}
	\caption{\textbf{Experimental setup for the MRR network measurements.} \textbf{(a)} Schematic of the setup employed to produce the MRR responses to the previously measured DAS reflection signals. \textbf{(b)} Example preprocessed signal (showing only 3 reflection signals in the considered sequence), containing traces from the DAS, used to modulate the laser beam injected into the MRR network. The backscattering signals obtained in the DAS measurements were randomly shuffled into one sequence, in which they are separated by a \SI{10}{\micro\second} offset-power plateau and a low-power break of the same length. \textbf{(c)} Scheme of the employed photonic integrated circuit \(8\times 8\) MRR network. The network comprises one input port (grating coupler) and 5 output ports.}
 \label{fig:setup}
 \end{figure*}

The amplitude reflection signals produced by the DAS measurements were preprocessed and experimentally reproduced via laser amplitude modulation and injected in the the MRR network response (the employed experimental setup is shown in Figure \ref{fig:setup} (a)). In particular, the obtained signal sequence (see Figure \ref{fig:setup} (b)) was uploaded to an arbitrary waveform generator (AWG, Spectrum model DN2.663-02, 12-bit resolution), which was used to modulate IR laser light (around 1550 nm wavelength, by a tunable Pure Photonics laser) at 200 MS/s in an optical fiber, employing a high-speed electro-optic modulator (iXblue model MXAN-LN-10). This setup allowed us to reproduce optical signals with approximately the same amplitude variations of the original fiber sensor output, mimicking the case where the DAS sensor was directly and optically connected to the employed MRR network. It should be stressed that reproducing optical phase variations was not necessary, since they are typically much slower than the travel time of optical signals through the PIC (tens of picoseconds), making the MRR network insensitive to them.
 
The modulated signal was then amplified (EDFA, Thorlabs) in order to achieve high enough optical signal power levels (up to around \SI{100}{\milli\watt}) for the nonlinear excitation of the MRR network. Then, a light polarization control stage was employed to set the appropriate TE (transverse electric) polarization for the excitation of the MRR network. Moreover, the input optical power was controlled via an electronic variable optical attenuator. Just before the insertion into the MRR network, a 1\% of the signal power was extracted and acquired by a slow photodetector (PD, 30 kHz bandwidth, New Focus model M2033), in order to monitor the average MRR network input power. Then, the resulting optical signal was fed into the MRR network via a cleaved optical fiber coupled to the MRR network input port (grating coupler, see scheme in Figure \ref{fig:setup} (c)). The output optical signals at the PIC output ports were extracted via another cleaved optical fiber, read out by a high-speed photodetector (\SI{600}{\mega\hertz}, Menlosystem model FPD610-FC-NIR), and digitized by an oscilloscope at 156 MS/s (Picoscope model 6000, 8-bit resolution).

We performed grid search measurements of the MRR response in a 3D parameter space, whose dimensions are given by: the output port of the MRR network to be used, the optical frequency of the laser (corresponding to a laser wavelength) and the input laser power. Namely, for each of the 5 output ports, we measured the network response to the input signal sequence at 20 different frequencies (192.66 THz, 192.67 THz, …, 192.85 THz). For each frequency value, we measured the MRR network response using 10 different laser power levels, corresponding to VOA power transmissions (10\%, 20\%, …,100\%). Moreover, for each employed AWG waveform, we also carried out a baseline measurement at a non resonant laser frequency and with maximum laser power (100\% VOA transmission). This allows us to obtain linear out of resonance baseline measurements where the output optical signal is not influenced by the MRR network, but is still affected by the non-idealities due to optical signal generation and acquisition (noise and limited vertical resolution). Considering the 17,500 reflection signals produced by the DAS measurements, in the MRR network measurement we acquired 17,517,500 MRR network responses to DAS reflection signals.

The integrated photonic circuit is composed by a MRR array based on silicon-on-insulator waveguides with a silicon core cross-section of \SI{450}{\nano\meter} \(\times\) \SI{220}{\nano\meter}, embedded in a silica cladding. The MRR have a racetrack shape with a bend radius of 7 µm and straight coupling sections of \SI{0.71}{\micro\meter}. The gap between the bus waveguides and the MRR waveguide is \SI{0.2}{\micro\meter} long. The distance between the centers of adjacent MRR on the same line is \SI{22.7}{\micro\meter}, while MRR on adjacent lines are horizontally displaced by \SI{11.35}{\micro\meter}. The PIC was fabricated by IMEC (Leuven, Belgium).

 \subsection{Down-sampling of the MRR network's output signals}
\label{methods_downsamp}
The optical signals at the output of the MRR network were digitized by the oscilloscope at a rate of 156 MS/s, while the original optical signals, consisiting in the backcattering from the FUT, were digitized at 200 MS/s. The signals in the latter were employed both to modulate the input of the MRR network and to estimate the baseline results, without passing through the experimental setup and processing pipeline for MRR network measurements. 

In order to down-sample an acquired signal to a target sampling rate (say, e.g., \SI{1}{\mega\hertz}), we apply (via software) a 1D Gaussian low-pass filter to adjust the signal's bandwidth to the target one (e.g., from \SI{200}{\mega\hertz} to \SI{1}{\mega\hertz}). Then, we apply a simple linear interpolation to obtain the corresponding sample rate (e.g., 1 MS/s). This down-sampling  operation approximates the employment of a photodetector with a lower bandwidth than that of its amplitude-modulated input signal, and a subsequent digitization process at a sampling rate corresponding to the photodetector bandwidth (e.g., 1 MS/s for \SI{1}{\mega\hertz} bandwidth).

\section*{Acknowledgments}
This work has received funding from the PRIN 2022 project entitled "Time REsolved multiparametric Sensing with opticAl Unstable REservoir - 
TRESAURE."(2022AEEKNC).
The authors would also like to thank Mattia Mancinelli, who started the writing of the proposal of the project,Peter Bienstman for useful discussions and for lending the photonic chip, as well as Isey Meka and Almaz Demisie for their valuable support in performing measurements using the DAS setup.

\section*{Data and code availability}
The raw data supporting the results of this study are available from the corresponding author upon reasonable request. The Python code employed to process the raw data will be uploaded to the public repository Zenodo once this work is published. In the meantime, the code can be provided by the corresponding author upon reasonable request.

\section*{Author contributions}
A.L., S.B., I.A. and L.P. conceived the integrated photonic experiment. Y.S.M. conceived and performed the DAS experiment. A.L. performed the integrated photonic measurements and the data processing. A.L. and Y.S.M. wrote the manuscript. All authors contributed to the revision of the manuscript. A.L., Y.S.M., I.A., S.B., C.J.O., F.D.P., L.P. supervised the work and contributed to useful discussions.

\section*{Competing interests}
The authors declare no competing interests.

\bibliographystyle{unsrt}  
\bibliography{references}

\end{document}